\input psfig.sty
\def\ptitlea{Semiclassical energy formulas for power-law and }
\def\ptitleb{log potentials in quantum mechanics}
\nopagenumbers 
\magnification=\magstep1 
\hsize 6.0 true in 
\hoffset 0.25 true in 
\emergencystretch=0.6 in 
\vfuzz 0.4 in 
\hfuzz 0.4 in 
\vglue 0.1true in 
\mathsurround=2pt 
\def\nl{\noindent} 
\def\nll{\hfil\break\noindent} 
\def\np{\hfil\vfil\break} 
\def\ppl#1{{\noindent\leftskip 9 cm #1\vskip 0 pt}} 
\def\title#1{\bigskip\noindent\bf #1 ~ \trr\smallskip} 
 
\def\Airy{{\rm Ai}}
\font\trr=cmr10    
\font\bf=cmbx10    
\font\sl=cmsl10    
\font\it=cmti10    
\font\trbig=cmbx10 scaled 1500 
\font\th=cmti10    
\font\tiny=cmr8    
\def\ng{>\kern -9pt|\kern 9pt} 
\def\hi#1#2{$#1$\kern -2pt-#2} 
\def\hy#1#2{#1-\kern -2pt$#2$} 

\def\sgn{{\rm sgn}} 
\def\half{{1 \over 2}} 
 
\def\nl{\noindent}    
\def\nll{\hfil\break} 
\def\dbox#1{\hbox{\vrule 
\vbox{\hrule \vskip #1\hbox{\hskip #1\vbox{\hsize=#1}\hskip #1}\vskip #1 
\hrule}\vrule}} 
 
\def\qed{\hfill \dbox{0.05true in}} 
\output={\shipout\vbox{\makeheadline\ifnum\the\pageno>1 {\hrule} \fi 
{\pagebody}\makefootline}\advancepageno} 
 
\headline{\noindent {\ifnum\the\pageno>1 
{\tiny \ptitlea \ptitleb\hfil page~\the\pageno}\fi}} 
\footline{} 
\newcount\zz \zz=0 
\newcount\q 
\newcount\qq \qq=0 
 
\def\pref #1#2#3#4#5{\frenchspacing \global \advance \q by 1 
\edef#1{\the\q}{\ifnum \zz=1 { %
\item{[\the\q]}{#2} {\bf #3},{ #4.}{~#5}\medskip} \fi}} 
 
\def\bref #1#2#3#4#5{\frenchspacing \global \advance \q by 1 
\edef#1{\the\q}{\ifnum \zz=1 { %
\item{[\the\q]}{#2}, {\it #3} {(#4).}{~#5}\medskip} \fi}} 
 
\def\gref #1#2{\frenchspacing \global \advance \q by 1 
\edef#1{\the\q}{\ifnum \zz=1 { %
\item{[\the\q]}{#2}\medskip} \fi}}

\def\sref #1{~[#1]} 
 
\def\references#1{\zz=#1 
\parskip=2pt plus 1pt 
{\ifnum \zz=1 {\noindent \bf References \medskip} \fi} \q=\qq 

\pref{\hallpn}{R. L. Hall and N. Saad, J. Math. Phys.}{38}{4909 (1997)}{}
\pref{\hallpow}{R. L. Hall, Phys. Rev. A}{39}{5500 (1989)}{}
\pref{\varma} {S. N Biswas, K. Datt, R. P. Saxena, P. K. Strivastava, and V. S. Varma, J. Math. Phys., No. 9}{14}{1190 (1972)}{}
\pref{\castro }{Francisco M., Ferndez and Eduardo A. Castro, Am. J. Phys., No. 10}{50}{921 (1982)}{}
\bref{\barn}{J. F. Barnes, H. J. Brascamp, and E. H. Lieb}{In: Studies in Mathematical Physics: Essays in Honor of Valentine Bargmann (Edited by E. H. Lieb, B. Simon, and A. S. Wightman)}{Princeton University Press, Princeton, 1976}{p 83}
\pref{\nieto }{ M. M. Nieto and L. M. Simons, Am. J. Phys., }{47}{634 (1979)}{}
\pref{\hio} {F. T. Hioe, Don MacMillen, and E. W. Montroll, J. of Math. Phys., No 7}{17}{(1976)}{}
\pref{\trus}{H. Turschner, J.Phys. A, No. 4}{12}{451 (1978)}{}
\pref{\hill}{B. J. B. Crowley and T. F. Hill, J. Phs. A, No. 9}{12}{223 (1979)}{}
\pref{\mark}{Mark S. Ashbaugh and John D. Morgan III, J. Phys. A}{14}{809 (1981)}{}
\pref{\reno}{R. E. Crandall and Mary Hall Reno, J. Math. Phys. }{23}{ 64 (1982) }{}
\pref{\vasan}{S.S.Vasan and M Seetharaman, J. Phys. A}{17}{2493 (1984)}{}
\pref{\seeth}{M. Seetharaman and S. S. Vasan, J. Phys. A}{18}{1041 (1985)}{}
\pref{\alhai}{A. D. Alhaidari, Int. J. Mod. Phys. A}{17}{4551 (2002)}{}
\pref{\cift}{H. Ciftci, E. Ateser, and H. Koru, J. Phys. A}{36}{3821 (2003)}{}
\pref{\loud}{R. Loudon, Am. J. Phys.}{27}{649 (1959)}{}
\pref{\hain}{L. K. Haines and D. H. Roberts, Am. J. Phys.}{37}{1145 (1969)}{}
\pref{\andrews}{M. Andrews, Am. J. Phys.}{44}{1064 (1976)}{}
\pref{\gesz}{F. Gesztesy, J. Phys. A}{13}{867 (1980)}{}
\pref{\boys}{L. J. Boys, M. Kmiecik, and A. Bohm, Phys. Rev. A}{37}{3567 (1988)}{}
\pref{\gord}{A. N. Gordeyev and S. C. Chhajlany, J. Phys. A}{30}{6893 (1997)}{}
\pref{\reys}{J. A. Reyes and M del Castillo-Mussot, J. Phys. A}{32}{2017 (1999)}{}
\bref{\reed}{M. Reed and B. Simon}{Methods of Modern Mathematical Physics IV: Analysis of Operators}{Academic, New York, 1978}{The min-max principle for the discrete spectrum is discussed on p75}
\bref{\movr}{H. Movromatis}{Exercises in Quantum Mechanics}{Kluwer, Dordrecht, 1991}{}
\pref{\comp}{R. L. Hall and Q. D. Katatbeh, J. Phys. A}{35}{8727 (2002)}{}
\pref{\hallenv}{R. L. Hall, J. Math. Phys.}{34}{2779 (1993)}{}
\pref{\hallxn}{R. L. Hall, Col. Math. J.}{24}{366 (1993)}{} 
\pref{\hallkp}{R. L. Hall, J. Math. Phys.}{25}{2078(1984)}{}
\pref{\hallint}{R. L. Hall, J. Phys. G}{26}{981 (2000)}{} 
\bref{\gelfand}{I. M. Gel'fand and S. V. Fomin}{ Calculus of Variations}{Prentic-Hall, Englewood Cliffs, NJ, 1963}{}
\pref{\hallmix}{R. L. Hall, J. Math. Phys.}{33}{1710 (1992)}{}
\pref{\hallmb}{R. L. Hall, Phys. Rev. A}{51}{3499 (1995)}{}
\pref{\hallrho}{R. L. Hall, W. Lucha, and F. F. Sch\"oberl, J. Math. Phys.}
{43}{1237 (2002)}{} 
\pref{\hallwsum}{R. L. Hall, W. Lucha, and F. F. Sch\"oberl, J. Math. Phys.}{43}{5913 (2002)}{[With some scale and figure corrections, math-ph/0110015; an Erratum is to appear in JMP]}

} 
 
\references{0} 
\topskip 2pt
\trr 
\ppl{CUQM-97}\ppl{math-ph/0305020}\medskip 

\vskip 0.6 true in 
\centerline{\trbig \ptitlea}
\vskip 0.2 true in
\centerline{\trbig \ptitleb}

\vskip 0.4true in
\baselineskip 12 true pt 
\centerline{\bf Richard L. Hall and Qutaibeh D. Katatbeh}\medskip
\centerline{\sl Department of Mathematics and Statistics,}
\centerline{\sl Concordia University,}
\centerline{\sl 1455 de Maisonneuve Boulevard West,}
\centerline{\sl Montr\'eal, Qu\'ebec, Canada H3G 1M8.}
\vskip 0.2 true in
\centerline{email:\sl~~rhall@mathstat.concordia.ca}
\bigskip\bigskip
\baselineskip = 18true pt 
 
\centerline{\bf Abstract}\medskip 
We study a single particle which obeys non-relativistic quantum mechanics in $\Re^N$ and has Hamiltonian $H=-\Delta+V(r),$ where $V(r) = \sgn(q)r^{q}.$   If $N\geq 2,$ then $q > -2,$ and if $N = 1,$ then $q > -1.$ The discrete eigenvalues $E_{n\ell}$ may be represented exactly by the semiclassical expression $E_{n\ell}(q)=\min_{r>0}\{P_{n\ell}(q)^2/r^2+ V(r)\}.$ 
The case $q = 0$ corresponds to $V(r) = \ln(r).$ By writing one power as a smooth transformation of another, and using envelope theory, it has earlier been proved that the $P_{n\ell}(q)$ functions are monotone increasing.  Recent refinements to the comparison theorem of QM in which comparison potentials can cross over, allow us to prove for $n = 1$ that $Q(q)=Z(q)P(q)$ is monotone increasing, even though the factor $Z(q)=(1+q/N)^{1/q}$ is monotone decreasing. Thus $P(q)$ cannot increase too slowly. This result yields some sharper estimates for power-potential eigenvlaues at the bottom of each angular-momentum subspace.\medskip

\nll PACS: 03.65.Ge 
\np 
\topskip 20pt
\title{1.~~Introduction} 

In this paper we study a certain representation, the \hi{P}{representation}, for the Schr\"odinger spectra generated by the power-law potentials $f(r)=\sgn(q)r^q $ in $N$ spatial dimensions. Considerable interest has been shown in the Schr\"odinger spectra generated  by this elementary class of potentials\sref{\hallpn-\cift}.  The Hamiltonian $H$ is given explicitly by 
$$H=-\Delta+v\ \sgn(q) r^q,\ {\rm where}\ r=\|{\bf r}\| \ {\rm and } \ v>0,~{\rm and}~q\ne 0, \eqno{(1.1a)} $$
\nl where $q > -1$ for $N = 1,$ and $q > -2$ for $N \geq 2.$ Corresponding to the case $q = 0$ we have
$$H=-\Delta+v\ln(r),\ v>0.\eqno{(1.1b)} $$
\nl It is certainly possible to include the $\log$ potential as a limiting case of the
power potentials if in place of the potential family $f(r) = \sgn(q)r^q,$ we use 
$V(r,q) = (r^q-1)/q$  whose limit as $q\rightarrow 0$ is $V(r,0) = \ln(r).$ However, we have chosen instead to leave the power-potentials themselves in their simplest form and incorporate the $q\rightarrow 0$ limit smoothly in the spectral domain by means of the \hi{P}{representation}. This limit will be discussed again in this section, after the \hi{P}{representation} has been introduced. As with Eq.(1), our policy of favouring simple powers will again lead to two equations instead of one at various points in the development. 
\par The operators $H$ have domains ${\cal D}(H)\subset L^2(R^N),$ they are bounded below, and essentially self adjoint. For the most part we shall be concerned with the cases $N \geq2,$ but we may also include $N = 1$ provided $q > -1.$ The one-dimensional hydrogen atom ($N =1,\ q = -1$) has
been extensively studied\sref{\loud-\reys} but requires special side conditions not consistent with the class of problems we consider in this paper.  For the operators we consider, the essential spectrum is in $[0,\infty)$ and, by using a normalized Gaussian trial function $\phi,$ it is easy to select a scale so that $(\phi, H\phi) < 0,$ thus estalishing the existence of a discrete eigenvalue; for $q > 0,$ the entire spectrum is discrete\sref{\reed}.  
\nl The eigenvalues $E^N_{n\ell}$ for the power-law potential can be labelled by two quantum numbers, the total angular momentum $\ell = 0,1,2,\dots,$ and a `radial' quantum number, $n = 1,2,3,\dots,$ which represents $1$ plus the number of nodes in the radial part of the wave function. These eigenvalues satisfy the relation   
 $E^N_{n\ell}\le E^N_{m\ell},\ n<m.$ With our labelling convention, the eigenvlaue $E^N_{n\ell}(q)$ in $N\geq 2$ spatial dimensions has degeneracy $1$ for $\ell=0$ and, for $\ell>0,$ the degeneracy is given\sref{\movr} by the function $\Lambda(N,\ell)$, where

$$\Lambda(N,\ell)=(2\ell+N-2)(\ell+N-3)!/\{\ell!(N-2)!\},\quad N \geq 2, \ \ell>0.\eqno{(1.2)} $$

 \par We first review some general elementary results for the power-law eigenvalues\sref{\hallpow}. Nieto and Simons\sref{\nieto} have proved that the eigenvalues $E_n=E^1_{n0}$ for the power-law potentials in one dimension increase with the quantum number $n$ at a higher rate when $q$ is greater. However, for any $q$, this increase never attains $n^2,$ i.e., $\lim_{n\rightarrow\infty}E_n/n^2=0,\  q<\infty.$  In general, the dependence of the eigenvalues $E^N_{n\ell}$ on the coupling parameter $v$ may be established with the aid of elementary scaling arguments in which $r$ is replaced by $\sigma r,$ where $\sigma >0$. We find that

 $$E^N_{n\ell}(v)=v^{2/(q+2)}E^N_{n\ell}(1).\eqno{(1.3)}$$

\nl Thus, without loss of generality, we may limit further discussion to the case of 
unit coupling, $v = 1.$  We shall henceforth let expression such as $E(q)$ represent the
dependence of an eigenvalue of unit coupling on the power $q.$

\par We do have {\it some} exactly solvable potentials in $N$ dimensions. For example, for the well-known hydrogenic atom and the harmonic oscillor potentials we have for $n = 1,2,3, ....$

$$E^N_{n\ell}(-1)=-[2(n+\ell+N/2-3/2)]^{-2},\quad N\geq 2,\eqno{(1.4)}$$
\nl and 
$$E^N_{n\ell}(2)=4n+2\ell+N-4,\quad N\geq 2,\eqno{(1.5a)}$$
\nl and in one dimension (keeping $n = 1,2,3, \dots)$
$$E_{n}(2)=2n-1,\quad N = 1.\eqno{(1.5b)}$$

\nl Analytical solutions are also possible for the linear potential in one dimension, and for the $S$ states in three dimensions. For $N=1\ {\rm and } \ N=3 $ the repulsive $1/r^2$ term in the `effective potential' $V_{\rm eff}(r)=(N-1)(N-3)/{4r^2},$ obtained using the transformation $\psi(r)=\phi(r)/r^{(N-1)/2}$, is zero. The exact solution in these cases is in terms of the zeros of Airy's function $\Airy(r)$ in three dimensions and the zeros of the first derivative $\Airy'(r)$ of Airy's function in one dimension.  We have 

$$E^1_{n}(v)=v^{2\over 3} r_{n+1},\quad \Airy'(-r_{n+1})=0,\ n=0,1,2,\dots .\eqno{(1.6)} $$

$$E^3_{n0}(v)=v^{2\over 3} r_n,\quad \Airy(-r_n)=0,\ n=1,2,3,\dots .\eqno{(1.7)} $$

\nl Unfortunately, for $N=2$ or $N>3$, and for higher angular momenta $\ell > 0$ generally, exact solutions are unavailable at this time. However, by using Theorem~1\sref{\comp, Theorem~2} we have for $N\geq 2$ the general 
correspondence $E_{n\ell}^N=E_{n0}^{N+2\ell}$. In Figure~1 we exhibit the graphs of the eigenvalues $E^3_{n0}(q)$ for $n=1\dots 6.$ In the limit $q\rightarrow \infty$ the problem is equivalent to an infinite square well with width $1$ in $N$ dimensions.  Thus we have $\lim_{q\rightarrow \infty}|E^3_{n0}(q)|=(n\pi)^2.$ For small values of $q,$ the $|E(q)|$ curves are asymptotically like $|E(q)| \sim C|q/2|^{q/2}$ and have infinite slopes in the limit$q\rightarrow 0$\sref{\hallpow,\hallenv,\hallxn}. 

\par The approach in the present paper is to study a representation for $E_{n\ell}(q)$ which is smoother and easier to approximate than the `raw' eigenvalues themselves.  We shall write many of our equations for the case $N \geq 2:$ they are also valid for $N = 1$ provided $q > -1.$ In both cases we keep the convention $n = 1,2,3, \dots.$ We have:
$$E^N_{n\ell}=\min_{r>0}\left \{\left (P^N_{n\ell}(q)\over r\right )^2 +\sgn(q) r^q \right\} ,\quad q > -2,\ q \neq 0,\eqno{(1.8a)}$$ 
\nl and
$$E^N_{n\ell}=\min_{r>0}\left \{\left (P^N_{n\ell}(0)\over r\right )^2 + \ln(r) \right\}.\eqno{(1.8b)}$$ 
  
\nl The form of this representation, in which the kinetic energy is represented by $P^2/r^2$ and the 
power-potential is represented by itself, is what leads us to use the term `semiclassical' in the title of the paper: the two parts of the quantum-mechanical problem are replaced by 
simple real functions of $r,$ scaling as the classical terms would scale, and their sum is exactly equal to the quantum-mechanical energy. This is a quite different use of the term `semiclassical' from that describing a reformulation of the quantum-mechanical problem itself. Such a method is
 the JWKB approximation which has been applied to estimate the pure-power spectra\sref{\vasan,\seeth} and could in principle therefore be employed to approximate $P^N_{n\ell}(q)$: however, this approach would not yield exact analytical information about the \hi{P}{functions}, such as bounds or convexity.  The existence of the representation $P(q)$ for $E(q)$ is guaranteed because the functions
$$g(P,q)= \min_{r>0}\left \{\left (P\over r\right )^2 +\sgn(q) r^q \right\}=\sgn(q)\left(1+{q\over 2}\right)\left({{2P^{2}}\over {|q|}}\right)^{q\over {2+q}} \quad q > -2,\ q \neq 0,\eqno{(1.9a)}$$
\nl and
$$g(P,0)= \min_{r>0}\left \{\left (P\over r\right )^2 + \ln(r) \right\}
 = \half(1+\ln(2)) + \ln(P).\eqno{(1.9b)}$$
\nl are monotone increasing in $P$. Indeed we find
$${\partial g\over \partial P}(P,q)= P^{{q}\over{2+q}}\left({|q|\over {2P} }\right)^{2\over q+2}>0,\quad 
q > -2,\ q\ne 0\eqno{(1.10a)}$$
\nl and
$${\partial g\over \partial P}(P,0)={1\over P} > 0.\eqno{(1.10b)}$$
\nl  From $(1.4)$ and $(1.5)$ we find:

$$P^N_{n\ell}(-1)=(n+\ell+N/2-3/2),\quad N \geq 2,\eqno{(1.11)}$$
\nl and 
$$P^N_{n\ell}(2)=(2n+\ell+N/2-2),\quad N\geq 2,\eqno{(1.12a) }$$
\nl and in one dimension (keeping $n = 1,2,3, \dots)$
$$P_{n}(2)=(n-\half),\quad N = 1.\eqno{(1.12b) }$$

\nl In Table~1 we exhibit some numerical values for $P^N_{n\ell}(1)$. The case $q = 0$ corresponds {\it exactly} to the $\ln(r)$ potential\sref{\hallenv}. In this paper we shall usually denote by $E(q)$ and $P(q)$ the ground-state eigenvalues and \hi{P}{functions} in $N$ dimensions. 

\par We now return briefly to the question of considering the $\log$ potential as the limit of the family
 $V(r,q) = (r^q-1)/q,$ as $q\rightarrow 0,$  where we define $V(r,0) = \ln(r).$ A useful feature of the \hi{P}{representation} is that, for a given eigenvalue, only one \hi{P}{number} is required to determine the eigenvalue ${\cal E}$ corresponding to the `scaled' power potential $A + B \sgn(q)r^q,\ B>0.$ Thus, we may write (exactly) 
$${\cal E}^N_{n\ell}(A,B,q)=\min_{r>0}\left \{\left (P^N_{n\ell}(q)\over r\right )^2 +A+B\sgn(q) r^q \right\} ,\quad q > -2,\ q \neq 0,\ B>0.\eqno{(1.13)}$$ 
\nl In particular, with $A = -1/q,\ B = 1/|q|$ we have
$$V(r,q) = (r^q-1)/q\quad\Rightarrow\quad{\cal E}^N_{n\ell}(q)=\min_{r>0}\left \{\left (P^N_{n\ell}(q)\over r\right )^2 +{{r^q -1}\over q} \right\} ,\quad q > -2,\ q \neq 0.\eqno{(1.14)}$$
\nl Provided $P(q)$ is continuous, it follows immediately from (1.14) that   
 $$V(r) = \ln(r)\quad\Rightarrow\quad{\cal E}^N_{n\ell}=\min_{r>0}\left \{\left (P^N_{n\ell}(0)\over r\right )^2 +\ln(r)\right\}.\eqno{(1.15)}$$
\nl As we mentioned above, the continuity (in fact, monotonicity) of $P^N_{n\ell}(q)$ was proved in Ref.\sref{\hallpow}.  It is our opinion that the advantage of accommodating this limit easily does not justify the concomitant complication of having to work, for example,
with a harmonic oscillator having the form $V(r,2) = (r^2-1)/2.$ 

\par For $N\geq 2,$ the \hi{P}{numbers} and the underlying eigenvalues $E_{n\ell}^{N}$ satisfy the relation $P_{n\ell}^N=P_{n0}^{N+2\ell}$. This result is obtained using the following theorem\medskip
\noindent {\bf Theorem 1.}\sref{\comp, Theorem 2}
{\th \noindent Suppose that $H = -\Delta +V(r),$ where $V(r)$ is a central potential in $N\geq 2$ dimensions, has a discrete eigenvalue $E_{n\ell}^{N},$ then  $E_{n\ell}^{N} = E_{n 0}^{N+2\ell}.$}
\nl This theorem expresses the invariance of the eigenvalues with respect to changes in $\ell$ and $N$ that leave the sum $N + 2\ell$ invariant. 
\par The advantage of the \hi{P}{representation} is illustrated by comparing Figure~1 with Figure~2 which show, 
respectively, the eigenvalues $E_{n\ell}(q)$ and the corresponding \hi{P}{representations} $P_{n\ell}(q)$ for the case $N = 3.$   The \hi{P}{functions} of Figure~2 are evidently monotone increasing. This property has been proved mathematically by means of envelope theory\sref{\hallpow}: one power $q$ was
written as a smooth transformation of another $p$, and then the limit $p\rightarrow q$ 
was taken in the \hi{P}{picture}. The infinite slopes of $E(q)$ at $q = 0,$ mentioned above,
 are not visible in 
Figure~1 because the approach of the slopes to infinity is very slow for such functions: if
, for example, we consider\sref{\hallxn} the function $f(q) = |q|^{q},$ then, although $f'(0) = -\infty,$ we have $f'(10^{-5}) \approx -10.51.$

\par The principal result of the present paper is Theorem~4, to the effect that for $N\geq 1,$ $Q(q)=Z(q)P(q)$ is monotone increasing, where $Z(q)=(1 + q/N)^{1\over q}:$
this result is stronger than the monotonicity of $P(q)$ because the factor $Z(p)$ is decreasing;
thus we know more about $P(q)$ than we did. This theorem is proved in Section~2 and principally concerns the power-law potentials, but also treats
the $\log$ case by the use of the limit $q\rightarrow0$ and continuity.
As consequences of Theorem~4 we shall be able to derive some specific formulas for upper and lower bounds for the power-law energy eigenvalues, by using nearby comparisons.
However, it should be clearly emphasized at this point that the main purpose of the present paper is to strengthen our knowledge of the monotone function $P(q).$ 

\par Theorem~4 has been made possible by the emergence of generalized comparison theorems that allow comparison potentials to cross over and still predict spectral ordering. 
In Section~2 we restate the generalized comparison theorem (Theorem~4, of Ref.\sref{\comp}) which becomes Theorem~2 here, and we state Theorem~3 (Theorem~7, of Ref.\sref{\comp}), which provides explicit sufficient conditions for the application of Theorem~3 under a variety of potential crossing schemes. Theorem~3 allows us to prove our main result, Theorem~4. In Section 3, we use Theorem~4 to prove Theorem~5 which sharpens the envelope bounds found earlier in Ref.\sref{\hallpow}.  The earlier result used `envelope theory' based on the `standard' comparison theorem, which may be written $V_1 < V_2 \Rightarrow E[V_1] < E[V_2].$ As an illustration of Theorem~5 we apply it to generate spectral bounds for the bottom of the spectrum of $-\Delta + r^{3\over 2}$ in dimensions $N = 3\dots 10.$

 \title{2.~~Power-law potentials and generalized comparison theorems }
We now discuss the generalized comparison theorems which we shall apply to obtain our main result. We consider the two eigenproblems $(-\Delta + V_1(r))\psi_1(r) = E[V_1]\psi_1(r)$ and 
$(-\Delta + V_2(r))\psi_2(r) = E[V_2]\psi_2(r)$ in $N\geq 1$ dimensions, where $\psi_i(r),\ i = 1,2,$ are the respective
ground states (or the bottoms of angular-momentum subspaces labelled by a fixed $\ell\geq 0$).
 
\noindent{\bf Theorem 2.}\sref{\comp, Theorem 4}
$$k(r)=\int_0^r[V_1(t)-V_2(t)]\psi_i(t)t^{N-1}dt < 0,~\forall r>0,~i=1~{\rm or}~2 \Rightarrow E[V_1] < E[V_2].\eqno{(2.1)}$$ 

\nll We stated this theorem (and the following theorem) with strict inequalites: the proofs are essentially the same as given in Ref.\sref{\comp}. It may be difficult to apply Theorem~2 in practice since the positivity of the function $k(r)$ depends on the detailed properties of the comparison potentials.  Thus it is helpful to have simpler sufficient conditions, depending on the number and nature of the crossings over of the two comparison potentials. In particular we shall employ the case of two crossings, and sufficient conditions not involving the wave function.  Thus we have:\par
\noindent{\bf Theorem 3.}\sref{\comp,~Theorem~7}
{\th If the potentials $V_1(r)$ and $V_2(r)$ cross  twice for $r>0$ at $r=r_1,~r_2$ $(r_1<r_2)$ with

\noindent(i) $V_1(r)<V_2(r)$ for $0<r<r_1$ and 

\noindent(ii) $\int_0^{r_2} [V_1(t)-V_2(t)]t^{N-1}dt=0 $
 
\noindent then,  
$$k(r)=\int_0^r[V_1(t)-V_2(t)]\psi_i(t)t^{N-1}dt < 0 ,~\forall r>0~,\ 
 i = 1\ {\rm or}\ 2, \eqno{(2.2)}$$ 
 \noindent from which $E[V_1] < E[V_2]$ follows, by Theorem 2.}

Now we shall use the generalized comparison theorems to prove the monotonicity of a new function $Q(q),$ which does not `vary' so much as the function $P(q).$ As a consequence we shall be able to derive specific formulas for upper and lower bounds for the power-law energy eigenvalues.  We are able to prove the following:\bigskip
\noindent {\bf Theorem 4.}
{\th $P(q)$ represents via (1.8) the bottom $E(q)$ of the spectrum of $H = -\Delta + \sgn(q)r^q$, where $q \neq 0,$ and $q > -2,$ in $N\geq 2$ dimensions (or $q > -1$ for $N=1$). Define $Q(q)=(1+q/N)^{1/q}P(q),$ 
and $Q(0) = \lim_{q\rightarrow 0} Q(q) = e^{1/N}P(0),$ then $Q(q)$ is monotone increasing for $N\geq 2,\ q > -2$ (or $N = 1,\ q>-1).$}\medskip
\nl{\bf Proof:} Let $p>q,$ $p,q > -2$ for $N \geq 2$ and $p,q > -1$ for $N=1.$ 
We shall first suppose $p\neq 0$ and $q\neq 0$. Our goal is to prove that $Q(p)>Q(q).$ Assume that $V_1(r)=A+B\ \sgn(p)r^p$ and $V_2(r)=\sgn(q)r^q.$ Now, we choose $A$ and $B$ so that the potentials $V_1(r)$ and $V_2(r)$ cross over exactly twice, as illustrated in Figure~3. Let $A_1$ and $B_1$ represent the absolute values of the areas between the potentials. We vary $A$ and  $B$ so that $A_1=B_1$. Then Theorem~3 implies $E[{V_1}]\le\ (\ge) E[{V_2}]$ depending, as $r$ increases from zero, on which potential lies beneath the other when they first differ. Without loss of generality, we will assume, in this sense, that $V_1$ starts above $V_2;$ this leads to an upper bound. Since $V_1(r)$ is designed to intersect $V_2(r)$ exactly twice, we shall have two equations to solve to provide sufficient conditions for a bound. 
$$V_1(R)=V_2(R)\Rightarrow  A+B~\sgn(p) R^p=~\sgn(q)R^q ~{\rm and }\eqno{(2.3a)}$$ 
$$\int_0^R[V_1(r)-V_2(r)]r^{N-1}dr=0 \Rightarrow A {R^N\over N}+ B~\sgn(p) {R^{p+N}\over p+N}-\sgn(q) {R^{q+N}\over q+N}=0,~\eqno{(2.3b)} $$
\nl where $R$ is the second potential intersection point. We let $t=R^{p/q}$ and, solve $(2.3{\rm a})$ and $(2.3{\rm b})$ for $A(t)$ and $B(t),$ to find

$$A(t)={\sgn(q)N t^{q^2/p}(p-q)\over p(q+N)} \eqno{(2.4)}$$

$$B(t)= {|q|(p+N)\over |p|(N+q)t^{q/p(p-q)}}. \eqno{(2.5)}$$

\nl Without loss of generality, we may consider only the case when $p\ {\rm and}\ q>0,$ since the proof of the other cases is exactly similar. 
\noindent Theorem $3$ thus implies that
$$\min_t\{A(t)+B(t)^{2\over(p+2)} E(p)\} > E(q) \eqno{(2.6)}$$
\noindent Optimizing the left side over $t$, we find the critical point as follows.  We define 
$$ F(t)=A(t)+(B(t))^{2\over p+2}E(p)={N t^{q^2/p}(p-q)\over p(q+N)} +\left({q (p+N)\over p(N+q)t^{q/p(p-q)}}\right)^{2\over p+2}E(p) \eqno{(2.7)}$$

\nl We now simplify the equation to find the critical point in terms of $p$ and $q.$ We define the following:
$$n=q^2/p,$$
$$m={q\over p}(p-q)\left({2\over 2+p}\right), $$
$$a_1=\left({N(p-q)\over p(q+N)}\right), $$
\nl and
$$b_1= \left({q (p+N)\over p(N+q)}\right)^{2\over 2+p} E(p).$$
\nl Thus we have
$$F(t)=a_1 t^n+b_1 t^{-m} $$

$$F'(t)=a_1 n t^{n-1}-b_1 mt^{-m-1} $$
 \nl for which the minimum occurs at ${\hat t}=\left[{b_1m\over a_1 n}\right]^{1\over n+m}.$
 Meanwhile, the minimum value $F({\hat t})$ is given by
$$F({\hat t})=a_1\left[{b_1m\over a_1 n}\right]^{n\over n+m} +b_1 \left[{b_1m\over a_1 n}\right]^{-{m\over {n+m}}} 
$$

$$=a_1^{{m\over n+m}} b_1^{n\over n+m}\left[{m\over n}\right]^{-{m\over {n+m}}}\left[{m\over n}+1\right]\quad \geq E(q).$$
\nl By substituting $F(\hat{t})$ and $E(p)$ given by (1.9) in (2.6), we find that
$$\left({N(p-q)\over p(q+N)}\right)^{2(p-q)\over p(q+2)}\left[\left({q(N+p)\over p(q+N)}\right)^{2\over 2+p} \left( {p+2\over 2}\right)\left( {2P(p)^2\over p}\right)^{p\over p+2}\right]^{q(2+p)\over p(q+2)}\times$$ $$\left[{q(p+2)\over 2(p-q)} \right]^{{2(p-q)\over p(q+2)}} \left[ {p(q+2)\over q(p+2)}\right] > \left({q+2\over 2}\right)\left({2P(q)^2\over q}\right)^{q\over 2+q}. \eqno{(2.8)}$$

\nl By simplifying this expression, we find eventually that $Q(q)=(1+q/N)^{1/q} P(q)$
is monotone increasing, that is to say
$$Q(p) > Q(q). \eqno{(2.9)}$$
\nl Now for $N\geq 2,$ $P(q)$ is continuous, $q > -2,$ (or for $N =1,\ q > -1$), and, if we define $Z(0)= \lim_{q\rightarrow 0}Z(q) = e^{1/N},$ then $Q(0) = Z(0)P(0).$ It follows immediately that $Q(q)= Z(q)P(q)$ is continuous and monotone increasing $q > -2$ (or for $N =1,\ q > -1$).\qed\smallskip

The three functions $P(q),$ $Z(q),$ and $Q(q)$ are illustrated for $N = 3$ in Figure~4:   Theorem~4 states that in all dimensions $N \geq 1,$ $Q(q)$ is a monotone increasing function of $q.$ 

 \title{3.~~Application}
By using the monotonicity of the function $Q(q)$, we now prove a special comparison theorem
(a corollary to Theorem~4) for the comparison of eigenvalues generated by power-law potentials.  

\noindent {\bf Theorem 5.}~~
{\th Consider the power-law potentials $V_i(r)=\sgn(q_i)r^{q_i},\ q_{i} > -2,$ ($q_{i} > -1,\ {\rm for}\ N = 1$), $i=1,2,$  where  $q_1<q_2.$ Let $Z(q)=(1+q/N)^{1/q},$ $Z(0) = \lim_{q\rightarrow}Z(q) = e^{1/N},$ $Q(q) = Z(q)P(q),$ and $g(P,q)$ be given by (1.9a) and (1.9b), then 

\noindent (i) $\displaystyle E[V_1]<{E^U_1=g(P(q_2),q_1) }$

\noindent (ii) $\displaystyle E[V_2]>{ E^L_1=g(P(q_1),q_2)} $

\noindent (iii) $\displaystyle  E[V_1]<{ E^U_2=
 g\left({Q(q_2)\over Z(q_1)},q_1\right) } < E^U_1$

\noindent (iv) $\displaystyle  E[V_2]>{E^L_2=
g\left({Q(q_1)\over Z(q_2)},q_2\right) } > E^L_1$
} \smallskip

\noindent {\bf Proof:} We first establish the upper bound (iii). We note that the function $Z(q)=(1+q/N)^{1/q} $ is decreasing. Thus $q_1<q_2$, implies $Z(q_2)<Z(q_1)$, and by using the monotonicity of the functions $P(q)$\sref{\hallpow} and $g(P,q),$ we may conclude that $P(q_1)<Z(q_2)P(q_2)/Z(q_1)= Q(q_2)/Z(q_1)<P(q_2)$, which, in turn, implies $E[V_1]<E^U_2<E^U_1$. This proves (i) and (iii).
\nl After a reversal of the inequalities, the proofs for the lower bounds (ii) and (iv) follow similarly.\qed

We note that Theorem~5 includes applications to the $\log$ potential. For example, if $q_1 = 0$
and $q_2 = q > 0,$ then we have from Theorem~5~(iv)
$$E(q) > \min_{r>0}\left \{{\left({Q(0)\over {Z(q) r}}\right)^2+\sgn(q)r^q }\right \},\quad q > 0.\eqno{(3.1)}$$
\title{Example: $V(r)=r^{3\over 2}$}

We illustrate Theorem~5 by applying it to the potential $V(r)=r^{3\over 2}$ in $N\geq 3$ dimensions.  We first use the linear and the harmonic oscillator problems to obtain upper and lower bounds by envelope theory. That is to say, we first use Eq.(1.9a) to give the envelope lower bound 
ELP given by $g(P(1),3/2),$ and the envelope upper bound EUP given by $g(P(2),3/2).$  Then we use Theorem~5~(iv) to generate the improved lower bound ELQ given by  $g(Q(1)/Z(3/2),3/2),$ 
and Theorem~5~(iii) to generate the improved upper bound EUQ given by $g(Q(2)/Z(3/2),3/2).$  These results are shown in Figure~5, along with accurate numerical data EX, for $N = 3\dots 10:$ they illustrate the 
improvement obtained in the approximation when $Q$ is used rather then $P$ in the semiclassical energy formulas.

 \title{4.~~Conclusion}

The eigenvalues $E(q)$ of $H = -\Delta + \sgn(q)r^q,$ $q > -2,\ q\neq 0,$ may be conveniently represented by the functions $P(q),$ which are known\sref{\hallpow} to be positive, continuous, and monotone increasing. In the proof of the earlier result, each \hi{q}{potential} was written as a smooth transformation of a \hi{p}{potential} with definite convexity, and then `envelope theory' was applied.  The envelope method, in turn, depends on the `standard' comparison theorem of quantum mechanics.  In the present paper we use a stronger comparison theorem, valid for node-free states in $N$ dimensions, and we are able thereby to learn more about $P(q)$ for the bottom of 
each angular-momentum subspace ($n = 1$). If $N >1$ and $\ell >0,$ 
we use the equivalence $E_{1\ell}^{N} = E_{1 0}^{2\ell+N}.$  We have shown for all these problems that $Q(q) = P(q)Z(q)$ is monotone increasing, where the factor $Z(q) = (1+q/N)^{1/q}$ is decreasing.  This immediately leads to some sharpened spectral inequalites concerning pairs of power-law Hamiltonians.

\par The $P(q)$ functions are important for an established general lower bound for potentials which are sums of powers. Thus if $V(r) = \sum_q a(q)\sgn(q)r^q + a(0)\ln(r),$ then we have\sref{\hallpn, \hallmix} for the bottom of each angular-momentum subspace in $N\geq 2$ dimensions:
$$E_{1\ell}^N \geq \min_{r>0}\left\{{1\over{r^2}} + 
\sum_{q}a(q)\ \sgn(q)\left(P_{1\ell}^N(q)r\right)^q\ + a(0)\ln\left(P_{1\ell}^{N}(0)r\right)\right\}.$$
\nl This formula, which is easily extended to smooth mixtures defined by an integral, is exact whenever the non-negative `weight' $a(q)$ is concentrated on a single term.  The lower bound is preserved if the \hi{P}{numbers} are replaced by lower bounds to them.  Thus any information concerning these fundamental numbers for the power-law potentials immediately has application to this general lower bound. These numbers have yielded useful energy bounds also for the many-body problem\sref{\hallmb}, and for relativistic problems\sref{\hallrho, \hallwsum}.   
\par In spite of the simplicity of the power-law potentials and the attractive scaling properties of the corresponding Schr\"odinger eigenvalues, general results concerning the unit-coupling eigenvalues $E(q)$ seem to be difficult to obtain.  One might expect that the results of the present paper would extend to all the excited states, but we know of no way at present to prove such general results. Even more ellusive seems to be a proof of the apparent concavity of all the $P(q)$ functions, some of which are illustrated for $N = 3$ in Figure~2. The establishment of concavity of $P(q)$ (or better, $Q(q)$) would immediately yield a large number of new spectral inequalities arising from the use of tangents and chords to the corresponding graphs. 
\title{Acknowledgment} 
  Partial financial support of this work under Grant No. GP3438 from the Natural 
Sciences and Engineering Research Council of Canada is gratefully acknowledged. 
\np 
\noindent {\bf Table 1}~~The `input' \hi{P}{values} $P^N_{n 0}(1)$  
used in the general formula (1.8), for $N=2,3,\dots,12$. The same data applies to $\ell > 0$ since, by Theorem~1, we have $P_{n\ell}^{N} = P_{n 0}^{N+2\ell}.$ 
 
\baselineskip=16 true pt 
\def\vr{\vrule height 12 true pt depth 6 true pt}
\def\vra{\vr\hfill} \def\vrb{\hfill &\vra} \def\vrc{\hfill & \vr\cr\hrule}
\def\vrq{\vr\quad} 

$$\vbox{\offinterlineskip
 \hrule
\settabs
\+ \vrq \kern 0.9true in &\vrq \kern 0.9true in &\vrq \kern 0.9true in &\vrq \kern 0.9true in &\vrq \kern 0.9true in&\vrq \kern 0.9true in&\vrq \kern 0.9true in&\vrq \kern 0.9true in&\vr\cr\hrule
\+ \vra $N$ \vrb  $n=1$\vrb $n=2$\vrb $n=3 $\vrb $n=4 $  \vrc
\+ \vra 2 \vrb 0.9348\vrb    2.8063\vrb    4.6249\vrb    6.4416\vrc
\+ \vra 3 \vrb 1.3761 \vrb   3.1813 \vrb   4.9926 \vrb   6.8051\vrc
\+ \vra  4 \vrb  1.8735\vrb    3.6657\vrb    5.4700\vrb    7.2783\vrc
\+ \vra  5\vrb  2.3719\vrb    4.1550\vrb    5.9530\vrb    7.7570\vrc
\+ \vra 6\vrb   2.8709\vrb    4.6472\vrb    6.4398\vrb    8.2396\vrc
\+ \vra 7\vrb   3.3702\vrb    5.1413\vrb    6.9291\vrb    8.7251\vrc
\+ \vra 8\vrb   3.8696\vrb    5.6367\vrb    7.4204\vrb    9.2129\vrc
\+ \vra 9\vrb   4.3692\vrb    6.1330\vrb    7.9130\vrb    9.7024\vrc
\+ \vra 10\vrb   4.8689\vrb    6.6299\vrb    8.4068\vrb   10.1932\vrc
\+ \vra 11 \vrb 5.3686\vrb    7.1274\vrb    8.9053\vrb   10.7453\vrc
\+ \vra 12\vrb   5.8684\vrb    7.6253 \vrb   9.4045\vrb   11.2744\vrc
}$$
\np
\vskip 0.4in

\references{1}     
  
\np 
\hbox{\vbox{\psfig{figure=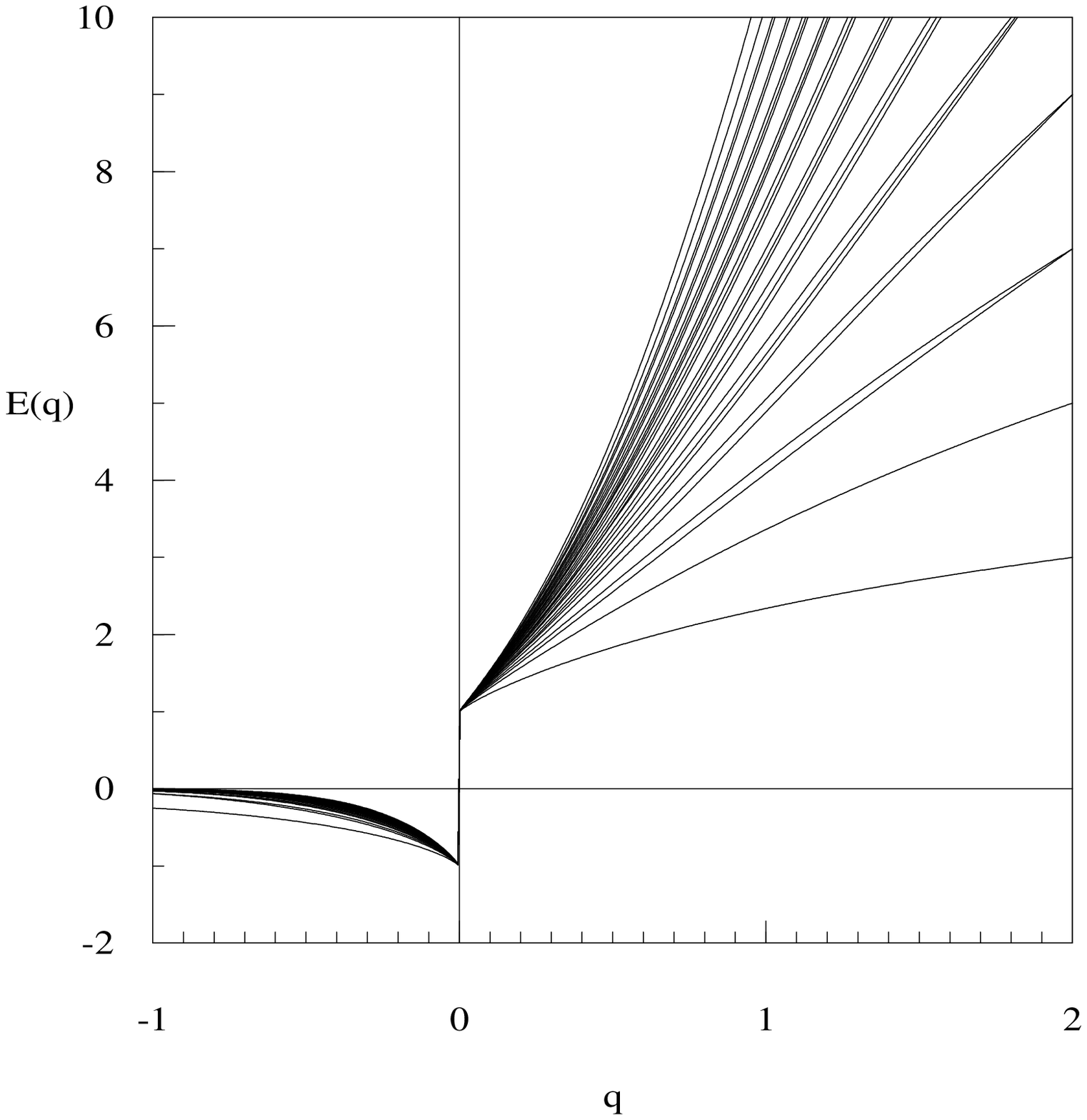,height=5in,width=4in,silent=}}}

\title{Figure 1.}
\nl The first 30 eigenvalues $E_{n\ell}(q),$ $1 \leq n \leq 5,$ $0 \leq \ell \leq 5,$ corresponding to the power potential $V(r) = \sgn(q) r^{q}$ in $N = 3$ dimensions. For $q > 0,$ the eigenvalues increase with $q$  from $1$ to $E_{n\ell}(2) = 4n + 2\ell - 1;$~  for $q < 0,$ they decrease (as $q$ increases) from $E_{n\ell}(-1) = -[2(n+\ell)]^{-2}.$ to $-1.$  Both sets of curves increase with $n$ and $\ell.$
\np
\hbox{\vbox{\psfig{figure=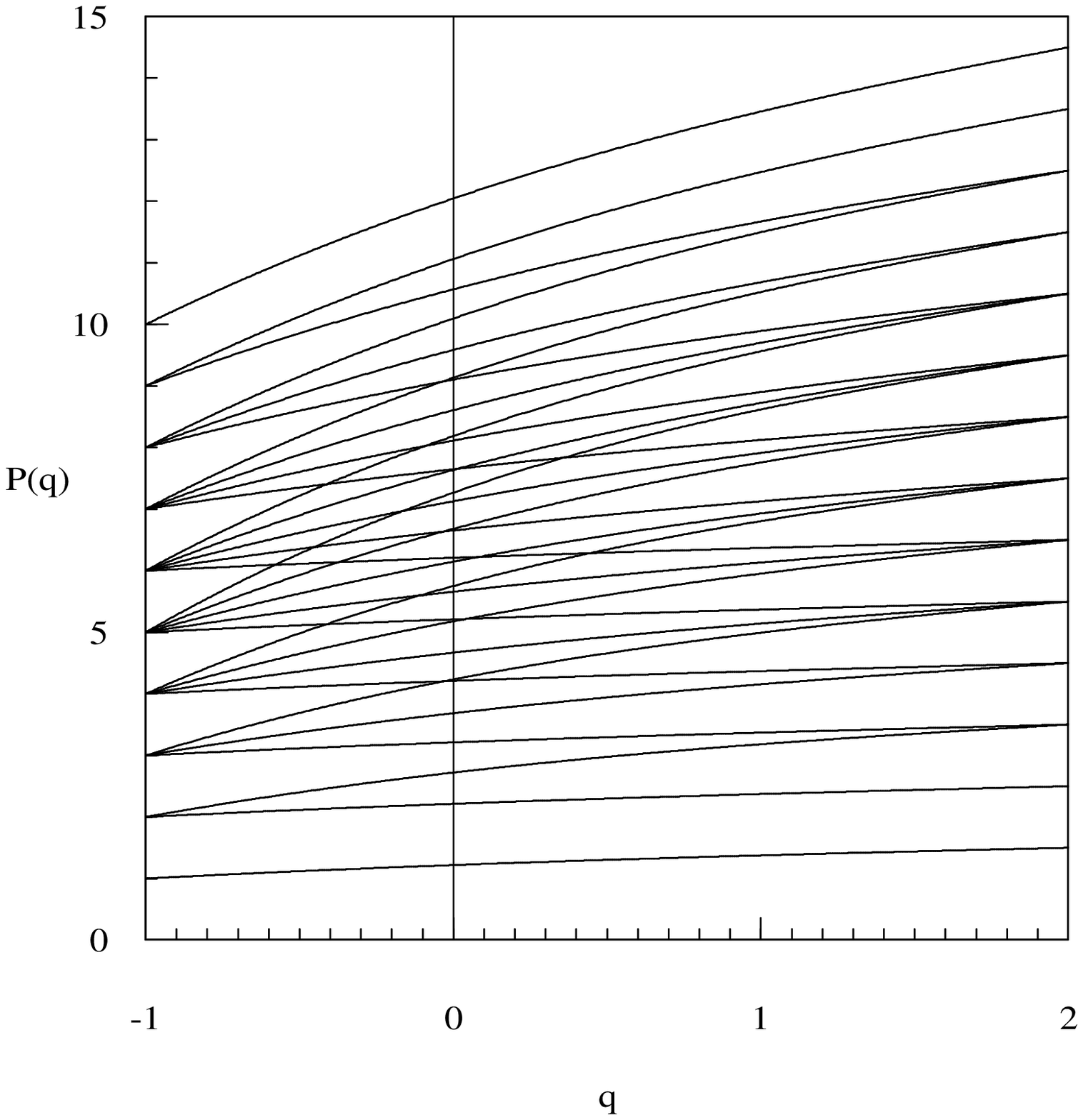,height=4in,width=4in,silent=}}}

\title{Figure 2.}
\nl In the $P$-representation, the same set of 30 eigenvalues shown in Fig.(1) now lie on monotone smooth curves.  The log-power theorem states that the $P$ values for the log potential are precisely $P_{n\ell}(0).$ As $q$ increases from $-1$ to $2$, the degeneracy of the Coulomb problem $P_{n\ell}(-1) = n + \ell$ evolves into the degeneracy of the harmonic oscillator $P_{n\ell}(2) = 2n + \ell - \half.$
\np
\hbox{\vbox{\psfig{figure=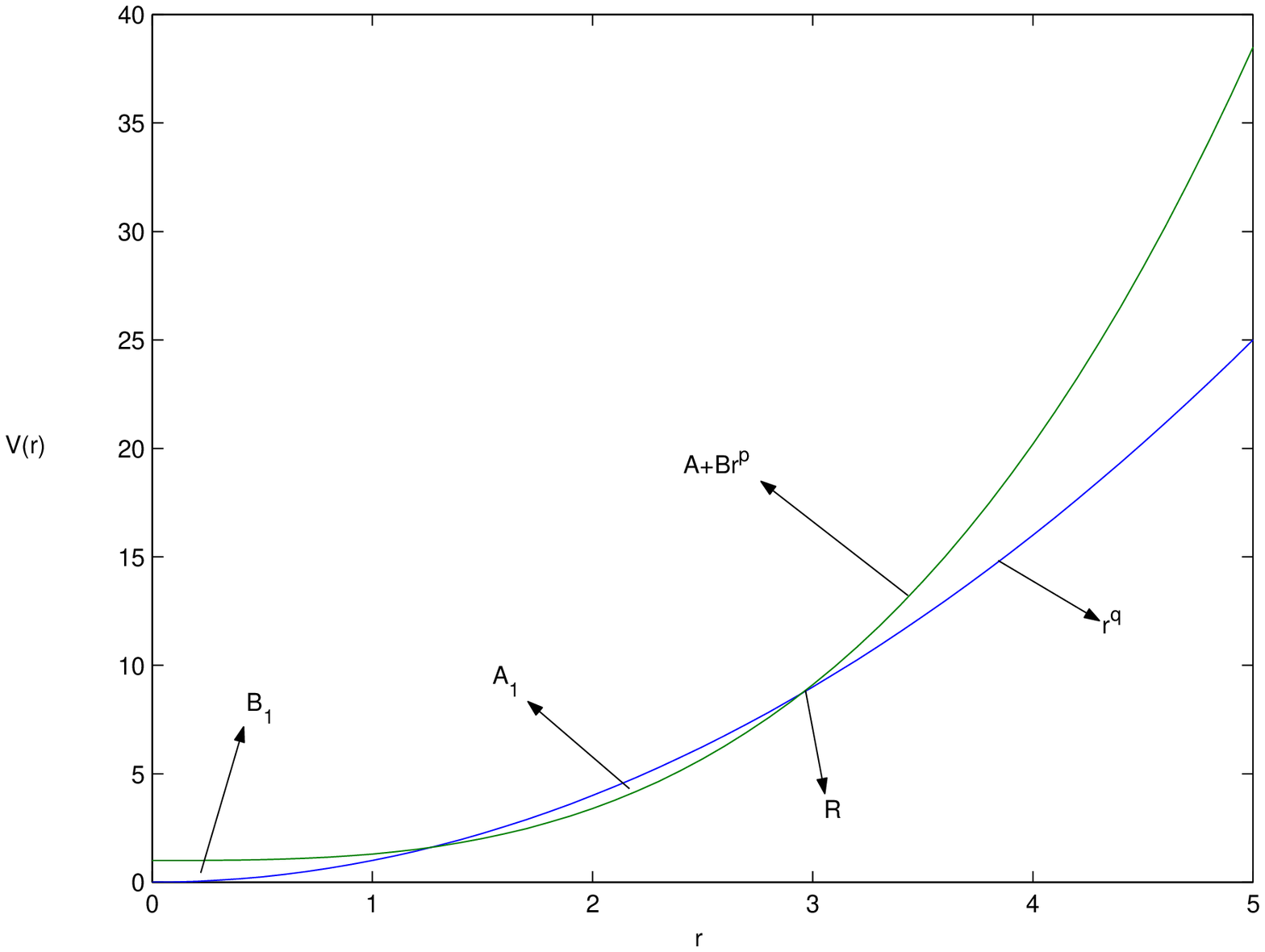,height=4.5in,width=5.5in,silent=}}}

\title{Figure 3.}
\nl The shifted linear potential $V_1(r)=A+Br^p$ used to estimate an upper bound for the eigenvalues corresponding to the potential $V_2(r)=r^q$. $A_1$ and $B_1$ are the absolute values of the inter-potential areas. We vary $A$ and $B$ so that $A_1=B_1$, where $R$ is the second intersection point. Thereafter, Theorem~3 implies that $E[V_2]\le E[V_1].$ This result is used to prove the monotonicity of $Q(q)$. 

\np
\hbox{\vbox{\psfig{figure=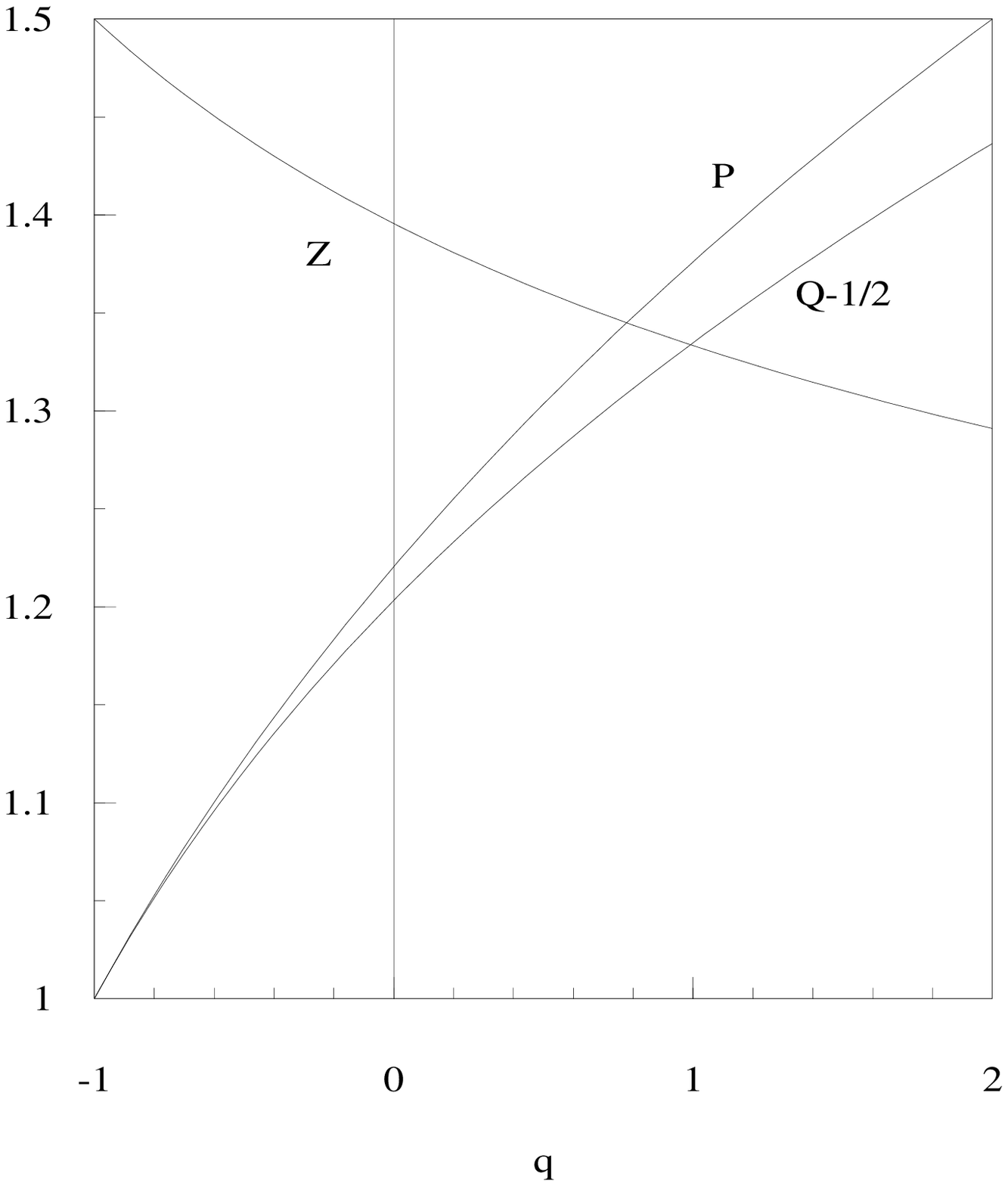,height=4in,width=4in,silent=}}}

\title{Figure 4.}
\nl The functions $P(q),\ Z(q),$ and $Q(q) = P(q)Z(q)$ for the ground state in 
dimension $N = 3.$  Theorem~4 states that for the ground state in all dimensions $N \geq 1,$ $Q(q)$ is monotone increasing with $q.$ 
\np
\hbox{\vbox{\psfig{figure=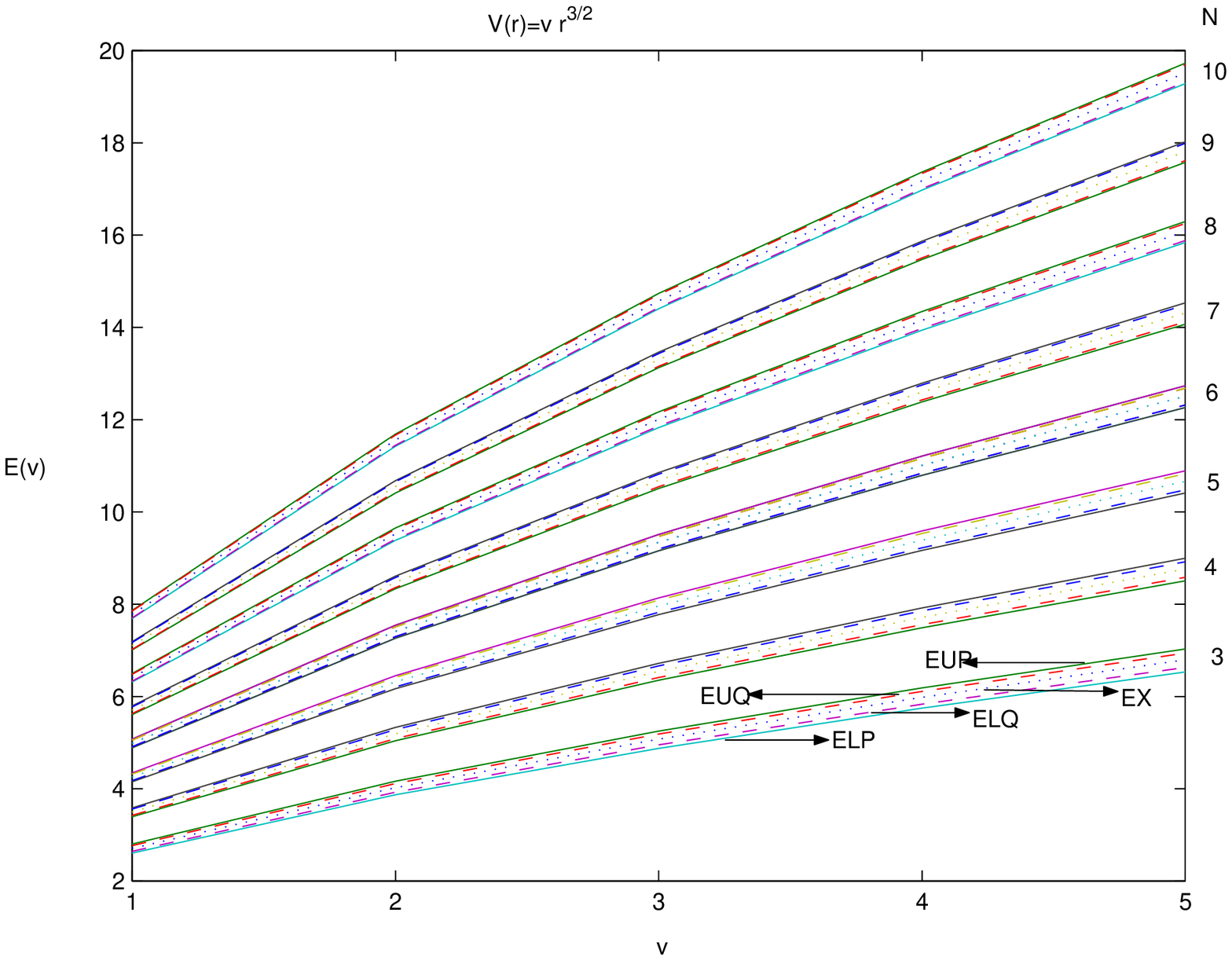,height=4.5in,width=5.5in,silent=}}}

\title{Figure 5.}
\nl Bounds on the eigenvalues $E^N_{10}(v)$ corresponding to the power potential $V(r)=vr^{3\over 2}$ in $N$ dimensions. The upper and lower bounds (full lines) are obtained by harmonic-oscillator tangents EUP, and linear tangents ELP (Theorem~5~(i),(ii)).  The dashed curves EUQ and ELQ represent respectively the improved upper and lower bounds (Theorem~5~(iii),(iv)). Accurate numerical data (dotted curves) EX are shown for comparison. 

 
\end